\documentclass{emulateapj}

\shorttitle{COSMIC-RAY ``ZEVATRON''}
\shortauthors{Honda \& Honda}

\begin{document}

\title{FILAMENTARY JETS AS A COSMIC-RAY ``ZEVATRON''}

\author{Mitsuru Honda\altaffilmark{1} and Yasuko S. Honda \altaffilmark{2}}

\altaffiltext{1}{Plasma Astrophysics Laboratory, Institute for Global Science, 
Mie 519-5203, Japan.}
\altaffiltext{2}{Department of Electrical and Information Engineering, Kinki 
University Technical College, Mie 519-4395, Japan; yasuko@ktc.ac.jp.}

\begin{abstract}
Strong, anisotropic turbulence reflecting magnetized filaments is
considered, to model the diffusive acceleration of particles by
shock waves in active galactic nucleus jets.
We address that at knot~A of the nearby M87 jet, the shock involving the
filamentary turbulence can accelerate an iron nucleus to zetta-eV
(ZeV; $10^{21}~{\rm eV}$) ranges.
A smaller value of the particle diffusion coefficient is found to be
essential to achieve a ZeV cosmic-ray accelerator, a ``Zevatron.''
\end{abstract}


\section{INTRODUCTION}

The morphology of self-organized filaments appears in a wide variety of
celestial objects, particularly, in splendid extragalactic jets
(e.g., Cyg~A: Perley et al. 1984; M87: Owen et al. 1989; 3C~273:
Lobanov \& Zensus 2001), which extend to kiloparsec to megaparsec scales
with narrow opening angles.
The underlying physics seems to be ubiquitous; thereby, we encounter the
similar, smaller scale structure in the Galactic center
(GC; Yusef-Zadeh et al. 1984; Yusef-Zadeh \& Morris 1987)
and even in laboratory plasmas (Tatarakis et al. 2003).
In particular, it is, at present, known that the GC region arranges
more than 80 linear filaments (Yusef-Zadeh et al. 2004), including the
nonthermal filaments, which probably embody the energetic flowing plasmas.
Ordinarily, such filaments are permeated by, more or less, ordered magnetic
fields, which participate in self-organizing the highly anisotropic structure.

Relating to the issues of cosmic-ray acceleration, these observational
results stimulate us to elaborate on a fundamental theory of particle
transport in the magnetized filaments, typical for astrophysical jets.
As for the diffusive shock acceleration (DSA) of particles
(e.g., Drury 1983), the acceleration rate depends largely on the
diffusion property, inferred from magnetic field strength,
configuration, and turbulent state around the shock.
According to the conventional resonant scattering theory, the charged
particles being bounded by the mean magnetic field (or magnetic flux tube)
are resonantly scattered by the fluctuating magnetic fields superposed on the
mean field; and in this context, the DSA scenarios of extremely high energy
(EHE) cosmic rays have been discussed for active galactic nucleus (AGN) jets,
considering the parallel (Biermann \& Strittmatter 1987;
Rachen \& Biermann 1993) and oblique (Honda \& Honda 2004)
mean fields with respect to the shock normal direction.
As the particles scatterer, the magnetohydrodynamic (MHD) turbulence of
Alfv\'en waves with a Kolmogorov spectrum was postulated, although the
actual turbulent state in the jets has been unresolved to date.

At this juncture, recent polarization measurements using very long baseline
interferometry began to reveal that some AGN jets are permeated by the
magnetic fields transverse to the jet axis (typically, 1803+784; Gabuzda 1999).
Over parsec scales, the fields are apparently so smooth that one cannot
understand their polarity solely by invoking a trail of the fast-mode
oblique shocks that refract the field lines to the off-axial direction.
The feasible scenario responsible for the observed results is to allow
huge currents launched from the central engine of the accretion disk
(Honda \& Honda 2002).
As is theoretically known, the uniform currents are catastrophically unstable
for the {\it electromagnetic} filamentation instability (Honda 2004 and
references therein) that breaks up a single beam into many filaments carrying
a limited current each (Honda 2000), and the nonlinear evolution results in
self-organizing toroidal magnetic fields (Honda et al. 2000a).
In fact, large-scale toroidal magnetic fields have recently been discovered
in the GC region (Novak et al. 2003), accompanied by many filaments.
We accordingly conjecture that the similar topology appears in
extragalactic objects (Medvedev \& Loeb 1999), inter alia, AGN jets.

In this Letter, we put forth the filamentary AGN jets as a
promising candidate for the cosmic-ray Zevatron (Blandford 2000).
From a generalized quasi-linear transport equation coupled with the spectral
intensity of the magnetized filamentary turbulence, we derive a diffusion
coefficient of relativistic particles, and install it in the DSA model,
in order to estimate the timescale of particle acceleration.
The remarkable feature is that, particularly for high-$Z$ nuclei, the
coefficient becomes smaller than that derived from the resonant scattering
theory invoking the MHD turbulence, leading favorably to shorter
acceleration time, viz., higher acceleration efficiency.
We here demonstrate that in a feasible parameter domain for the brightest
(radio to X-ray) knot, knot~A (e.g., Wilson \& Yang 2002), of the M87 jet,
an iron nucleus can be accelerated to ZeV ranges.

\section{KINETIC THEORY OF PARTICLE DIFFUSION IN\\*
MAGNETIZED CURRENT FILAMENTS}

\subsection{\it The Filamentary Jet Model}
Let us begin with Figure~1, illustrating a schematic view of our model for
AGN jets as an EHE cosmic-ray accelerator.
As now seems more likely, a jet, envisaged as one or several filaments
within the current resolution (Owen et al. 1989; Asada et al. 2000;
Lobanov \& Zensus 2001), is hypothesized to be {\it a bundle of
numerous filaments}, whose radial size is self-adjusted by an
effective Debye sheath (Honda et al. 2000a; Honda \& Honda 2002).
For finite ion abundance, the charge conservation law requires that
the filamentary currents must carry the net charges corresponding to the
number of electrons subtracted by that of positrons.
In addition, there is direct observational evidence that the jet transports
energy from the central engine to the radio lobe (Biretta et al. 1995).
Thus, the hot currents carried by the negatively charged electron-positron
fluids flow in the opposite direction, as shown in Figure~1, while
the cold return currents compensating for positive ionic charges flow in the
same direction via the filaments and/or external media (Honda \& Honda 2002).

\begin{figure}
\centerline{\includegraphics*[bb=40.0 0.0 660.0 520.0,
width=\columnwidth]{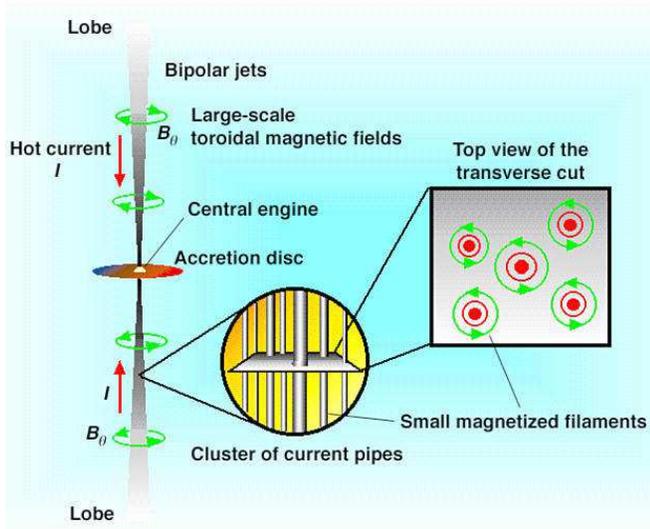}}
\vspace{0.2cm}
\caption{Schematic of AGN jets, which operate as ${\rm ZeV}$ particle
accelerators.
The hierarchical structure of filaments illustrates that a jet,
carrying huge current $I$, consists of numerous magnetized filaments
with the smaller radial sizes (Honda \& Honda 2002).
In the transverse section of sample filaments, the ``fish eyes''
symbolically represent the hot currents pointing out of the page,
and the loops with arrows the magnetic field lines.
In the envelope region of bundled filaments, the large-scale toroidal field
$B_{\theta}$ can appear, establishing a magnetotail.}
\end{figure}

From a macroscopic point of view, the vectors of the self-generated
magnetic field appear to be randomly oriented in the transverse direction
of the bundled filaments, such that their average is negligible,
except for the envelope region.
As far as the return currents flow, in part, via the external media (e.g.,
radio cocoons), the uncancelled magnetic field pervades the exterior of the
jet, establishing a long-range ``magnetotail'' of the toroidal field,
in accordance with a pioneering observation that revealed not-so-weak
(rather strong), ordered fields in the central kiloparsec-scale ``hole''
(cavity) of the inner radio lobe containing a jet (M87; Owen et al. 1990).
The toroidal field will play a significant role in radially confining the jet
(Honda \& Honda 2002), as is consistent with the observational fact that
pressure inside the jets is larger than that of the external media
(Perley et al. 1984; Owen et al. 1989).

\subsection{\it The Off-resonant Scattering of Particles by the
Magnetized Filaments}
In a forest of the magnetized filaments, diffusion property of injected
energetic particles is markedly different from that stemming from the
small-angle resonant scattering (Drury 1983).
When considering the DSA of the particles, it is necessary to know the
effective diffusion coefficient for the direction normal to the shock front,
$\kappa_{\rm n}$.
As outlined below, it can be derived from the Vlasov equation that
describes the collisionless scattering of relativistic test particles
by the fluctuating magnetic fields ${\bf B}$.
For simplicity, the fields are supposed to be purely transverse to the
$z$-axis, ${\bf B}={\bf B}_{\perp}({\bf r},t)$, where ${\bf r}=(x,y)$,
and electric fields are ignored, since they are preferably shorten out
in the propagation time of jets (Honda et al. 2000a).
Hereafter, the perpendicular and parallel subscripts refer to the
$z$-axis reflecting the direction of the current filaments.
We decompose the momentum distribution function of the particles
into the averaged and fluctuating part,
$f_{\bf p}=\left<f_{\bf p}\right>+\delta f_{\bf p}$,
and combine the averaged Vlasov equation,
$\left<{\rm D}f_{\bf p}/{\rm D}t\right>=0$, with the evolution
equation of $\delta f_{\bf p}$ that contains the term involving
$\left<f_{\bf p}\right>$, linear to ${\bf B}_{\perp}$.
For the manipulations, we carry out the Fourier transformation of all
fluctuating quantities in terms of both ${\bf r}$ and $t$:
${\cal F}({\bf r},t)=\int{\rm d}^{2}{\bf k}{\rm d}\omega
{\cal F}_{{\bf k},\omega}e^{i({\bf k}\cdot{\bf r}-\omega t)}$,
where ${\bf k}\perp{\hat {\bf z}}$.
As a result, we obtain
\begin{eqnarray}
{{{\rm d}\left<f_{\bf p}\right>}\over{{\rm d}t}}&=&-i{q^2\over c^2}
\left<\int{\rm d}^{2}{\bf k}{\rm d}^{2}{\bf k}^{\prime}
{\rm d}\omega{\rm d}\omega^{\prime}
e^{i\left[\left({\bf k}+{\bf k}^{\prime}\right)\cdot{\bf r}
-\left(\omega+\omega^{\prime}\right)t\right]}\right. \nonumber \\
&\times& \left. \left[{\bf v}\times\left({\bf k}^{\prime}\times
{\bf A}_{{\bf k}^{\prime},\omega^{\prime}}\right)\right]\right. \nonumber \\
&\cdot& \left. {\partial\over{\partial{\bf p}}}
\left\{{{\left[{\bf v}\times\left({\bf k}\times
{\bf A}_{{\bf k},\omega}\right)\right]}\over
{\omega-\left({\bf k}\cdot{\bf v}\right)}}
\cdot{\partial\over{\partial{\bf p}}}
\right\}\left< f_{\bf p}\right>\right>,
\end{eqnarray}
where ${\rm d}/{\rm d}t\equiv\partial/\partial t+
{\bf v}\cdot(\partial/\partial{\bf r})$,
${\bf A}_{{\bf k}^{(\prime)},\omega^{(\prime)}}=
A_{{\bf k}^{(\prime)},\omega^{(\prime)}}{\hat{\bf z}}$ denote the
Fourier components of the vector potential that conforms to
${\bf B}_{\perp}({\bf r},t)=\nabla\times A_{\parallel}({\bf r},t){\bf\hat z}$,
$c$ is the speed of light, and $q$ and ${\bf v}$ are the charge and
velocity of the particle, respectively.

Even for the strong nonlinear fluctuations,
the correlation function may be written as
$\left<A_{{\bf k},\omega}A_{{\bf k}^{\prime},\omega^{\prime}}\right>=
|A|^{2}_{{\bf k},\omega}\delta({\bf k}+{\bf k}^{\prime})
\delta(\omega+\omega^{\prime})$, where the Dirac $\delta$-function
has been introduced (Tsytovich \& ter~Haar 1995).
Assuming the stationary and homogeneous fluctuations, i.e.,
$\omega=-\omega^{\prime}$ and ${\bf k}=-{\bf k}^{\prime}$, and invoking the
causality principle to handle the resonant dominator, we can express
equation~(1) in the form of the generalized quasi-linear equation.
Especially for the typical current filamentation instabilities that include
the Weibel instability, the unstable mode is often quasi-static
(Kazimura et al. 1998; Medvedev \& Loeb 1999; Honda 2004) and isotropic
on the transverse two-dimensional plane (Montgomery \& Liu 1979), to give
$|A|_{{\bf k},\omega}^{2}\sim 2|A|_{k}^{2}\delta\left(\omega\right)$,
where $k=|{\bf k}|$.
Along these, we impose the {\it off-resonance} condition for wave-particle
interaction, $|{\bf k}\cdot {\bf v}|\gg\omega$, and take an average
over the pitch angle.
Then, for example, the integral [eq.~(1)] including the double partial
derivatives of $\partial/\partial p_{\parallel}$ reduces to
\begin{equation}
\left({{{\rm d}\left<f_{\bf p}\right>}\over{{\rm d}t}}\right)_{\parallel}
\sim{{16\pi q^{2}}\over c^{2}}v_{\perp}
{\partial^2\over{\partial p_{\parallel}^{2}}}\left<f_{\bf p}\right>
\int_{k_{\rm min}}^{\infty}{{{\rm d}k}\over k}I_{k},
\end{equation}
where $v_{\perp}=|{\bf v}_{\perp}|$ and $p_{\parallel}=|{\bf p}_{\parallel}|$.
Here we have defined the turbulent spectral intensity as
$I_{k}\equiv 2\pi k\left(k^{2}|A|_{k}^{2}/4\pi\right)$,
such that the magnetic energy density can be evaluated by
$u_{\rm m}\equiv B^2/(8\pi)=\int_{k_{\rm min}}^{\infty}I_{k}{\rm d}k$,
where $B^{2}\equiv\left<{\bf B}_{\perp}^2({\bf r})\right>$.
The minimum wavenumber $k_{\rm min}$ is set to $\pi/R$, where $R$ stands for
the radius of a bundle of the filaments, namely, the ``radius of the jet.''

The energy density of the fluctuations, $u_{\rm m}$, likely
becomes comparable to thermal pressure (Honda et al. 2000a).
In this sense, the anisotropic filaments may be regarded as
the ``{\it strong turbulence}.''
We see that the spectrum exhibits a power law, $I_{k}\propto k^{-\alpha}$,
with its index around $\alpha\sim 2$ (Montgomery \& Liu 1979), somewhat
larger than $\alpha_{\rm MHD}\approx 1-(5/3)$ for the classical MHD turbulence.
The kinetic simulations also indicated that the filament coalescence
led to the accumulation of larger filaments, releasing in part the free
energy of flows (Honda et al. 2000b).
These are compatible with the observed trends of the steepening of
filamentary turbulent spectra (Carilli \& Barthel 1996) and
the merging of filaments (Owen et al. 1989).

\subsection{\it The Diffusion Coefficient}
For a simple estimation, let us assign the EHE particles having their
energy of $E=cp$ and the isotropic momentum distribution
$\left<f_{p}\right>\propto p^{-\beta}\propto E^{-\beta}$ with $\beta\sim 3$
(Stecker \& Salamon 1999; de~Marco et al. 2003), so as to arrange
equation~(2) in the form of
$({\rm d}\left<f_{p}\right>/{\rm d}t)_{\parallel}=
\nu_{{\rm eff},\parallel}\left<f_{p}\right>$, where
$\nu_{{\rm eff},\parallel}$ reflects an effective collision frequency.
The spatial diffusion coefficient can be then estimated as
$\kappa_{\parallel}\sim v_{\parallel}^2/(2\nu_{{\rm eff},\parallel})
\sim (\sqrt{6}\pi/4)cE^{2}/[\beta(\beta-1)q^{2}B^2R]$, where we have used
$v_{\parallel}^2\sim v_{\perp}^2/2\sim c^2/3$ and the definition of
$u_{\rm m}$.
Furthermore, supposing the shock to propagate along the filaments ($z$-axis)
gives $\kappa_{\rm n}=\kappa_{\parallel}$.

Here it may be instructive to look at the ratio of $\kappa_{\rm n}$
to a conventionally used coefficient for $\alpha_{\rm MHD}=5/3$, i.e.,
the Kolmogorov MHD turbulence (Biermann \& Strittmatter 1987).
For convenience, we compare the mean field strength of the turbulent
magnetic fields in the MHD context to $B$
[$=\left(\left<{\bf B}_{\perp}^2\right>\right)^{1/2}$]
and introduce $q=Ze$, where $Z$ and $e$ are the charge number of the
particle and the elementary charge, respectively.
Then we obtain the ratio of
$\kappa_{\rm n}/\kappa_{\rm n,MHD}\lesssim 0.1[(E/1~{\rm ZeV})(26/Z)
(1~{\rm mG}/B)(100~{\rm pc}/R)]^{5/3}$.
It is found that for kiloparsec-scale jets, larger values of $Z$ and $B$
lead to a smaller $\kappa_{\rm n}$, thereby, favorably to a shorter
DSA timescale of ${\rm ZeV}$ particles (see \S~3).
Importantly, the gyroradius of injected energetic particles cannot be well
defined in the present model, because of
$|\left<{\bf B}_{\perp}\right>|\ll B$ in the interior of the jet (\S~2.1).
When considering the particle confinement, therefore, we need to compare the 
three-dimensional rms deflection of the accelerated particle to the
system size, rather than its gyroradius.
Relating to this, we find $\kappa_{\perp}/\kappa_{\parallel}\sim{\cal O}(1)$,
in contrast to a larger anisotropy that appears in the simple MHD.
It follows that the radial size, which is smaller than the length of jet,
affects the confinement.
Note here that the radially decaying magnetotail may play an additional role in
confining the leaky energetic particles with their long mean free path (mfp)
of $\lambda_{\perp}(\sim v_{\perp}/\nu_{{\rm eff},\perp})~\sim R$.

\section{ZeV ACCELERATION OF THE IRON NUCLEUS IN THE M87 JET}

We are concerned with nearby radio galaxy M87 (Virgo~A)
residing in the center of the Virgo Cluster, which is known as a
sub-TeV (possibly TeV) gamma-ray emitter (Aharonian et al. 2003).
For application to the jet in the core of the giant elliptical galaxy,
one issue worth noting is that a condensation of heavy elements was
discovered in the core (Gastaldello \& Molendi 2002).
Presuming that the abundance of heavy elements including iron
is finite in the jet, as confirmed in a Galactic microquasar jet
(Kotani et al. 1996), as well as referring to up-to-date observational
results around $\sim 100~{\rm EeV}$ ($10^{20}~{\rm eV}$), compatible with
an assumption of heavier composition primaries (Ave et al. 2000; Risse et al.
2004), we pay attention to the acceleration of iron nuclei (and protons).

Taking account of the Fermi type I mechanism that is plausible for the
knotlike regions of the M87 jet (Heinz \& Begelman 1997),
the mean acceleration time can be expressed as
$t_{\rm acc}\simeq [3/(U_{\rm 1}-U_{\rm 2})]
[(\kappa_{\rm n,1}/U_{\rm 1})+(\kappa_{\rm n,2}/U_{\rm 2})]$,
where the subscripts $i=1,2$ indicate the upstream and downstream region of
the shock and $U_{i}$ are the flow speeds in the shock rest frame
(Drury 1983).
When assuming the values of $\alpha$, $\beta$, and $R$ are constants,
we have $\kappa_{\rm n,1}=\kappa_{\rm n,2}$, because of the relation of
$B_{1}=B_{2}$ derived from the condition that the current density
$J_{\parallel}{\hat{\bf z}}$ must be continuous across the shock front.
Then the above expression of the acceleration time reduces, in the strong
shock limit (Biretta et al. 1991), to
$t_{\rm acc}\simeq 20\kappa_{\rm n}/U^{2}$, where
$\kappa_{\rm n}\equiv \kappa_{{\rm n},i}$, and
$U\equiv U_{1}$ is the shock speed in the laboratory frame.

Concerning the energy constraints relevant to the temporal scale, the
shortest timescale that most severely restricts the acceleration of
nuclei is arguably the shock propagation time, which is estimated as
$t_{\rm sh}\sim L/U$, where $L$ represents the propagation distance of the
shock.
It is noted that for the kiloparsec-scale jet, the synchrotron loss of the
accelerated nuclei is ignorable for the inherent weak $B$-field, and radial
adiabatic expansion as well is ineffective for the self-collimating jet
(Honda \& Honda 2002, 2004).
For the moment, balancing $t_{\rm acc}$ with $t_{\rm sh}$ is adequate for
estimating the achievable maximum energy, and then, solving for $E$ yields
the scaling of $E_{\rm m,t}\propto ZB(LRU)^{1/2}$ (for the temporal limit).
In addition, the particle confinement radius, $R_{\rm c}$, limits the
acceleration in terms of the spatial scale.
As remarked in \S~2.3, we expect $R_{\rm c}\gtrsim R$,
taking the effects of the magnetotail into consideration.
By simply equating the transverse mfp $\lambda_{\perp}$ with $R_{\rm c}$,
we obtain another scaling of the achievable maximum energy,
$E_{\rm m,s}\propto ZB{\tilde\rho}^{1/2}R$ (for the spatial limit),
where ${\tilde\rho}\equiv R_{\rm c}/R$.
As a consequence, in the optically thin regions distant from the central
engine, the actual maximum energy of accelerated nuclei can be expressed as
$E_{\rm m}={\rm min}(E_{\rm m,s},E_{\rm m,t})$.
Note that the larger $R$ leads to the higher $E_{\rm m}$, because the
accumulating larger filaments more largely contribute to the deflection of
particles.
This $R$ dependence is quite contrary to that deduced from the MHD model,
in which a larger resonant gyroradius involves a longer coherence time,
that is, longer acceleration time (Honda \& Honda 2004).

\begin{figure}
\centerline{\includegraphics*[bb=60.0 90.0 680.0 530.0,
width=\columnwidth]{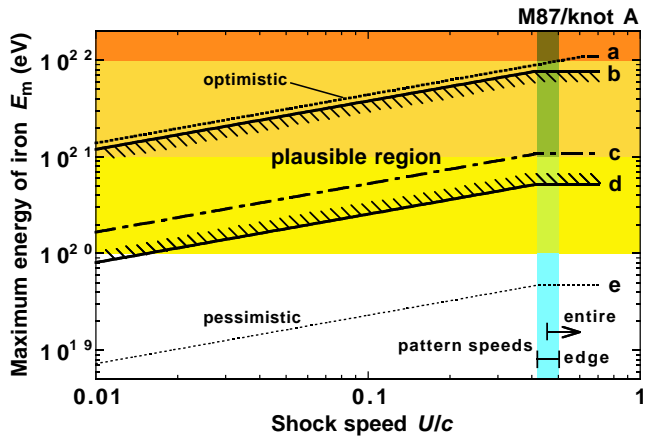}}
\caption{Maximum energy of the accelerated iron nucleus $E_{\rm m}|_{Z=26}$
vs. shock speed $U$ in units of $c$, for the M87 jet/knot~A.
The rms strength of the random magnetic fields, $B$, is set to
({\it a}, {\it b}) $4.59~{\rm mG}$ (Owen et al. 1989),
({\it c}) $640~{\rm\mu G}$ (Meisenheimer et al. 1996),
({\it d}) $310~{\rm\mu G}$ (Owen et al. 1989),
and ({\it e}) $28~{\rm\mu G}$ (Heinz \& Begelman 1997).
The viewing angle $\theta$ and the radial extension factor ${\tilde\rho}$ are
chosen as ({\it a}) $30^{\circ}$ (Bicknell \& Begelman 1996) and $2$, or
({\it b}$-${\it e}) $42.5^{\circ}$ (Biretta et al. 1995) and $1$, respectively.
Here the opening angle $\phi$ and the radius $R$ have been fixed to
$6.9^{\circ}$ (Reid et al. 1989) and $55~{\rm pc}$ (Biretta et al. 1991),
respectively.
The hatched area indicates $E_{\rm m}$ in a plausible parameter domain
[between ({\it b}) and ({\it d})].
For comparison, we also indicate the pattern speeds of the bright edge
of the knot ($0.42-0.50$; a band with a fat bar, where the darkly shaded zone
suggests the most plausible $E_{\rm m}$) and the entire knot
($>0.454$; arrow; Biretta et al. 1995).\\}
\end{figure}

For the evaluation of $E_{\rm m,t}$, we unfix the value of $U$, since
it seems to be uncertain yet, although the proper motion of some knots
was surveyed in detail and found to exhibit average speeds around $0.5c$
(Biretta et al. 1995).
An upper limit of $L$ is estimated by using an ad hoc relation,
$L/R\sim 360/\pi\phi\sin\theta$, for the opening angle
$\phi\approx 6.9^{\circ}$ (Reid et al. 1989)
and viewing angles $\theta\approx 42.5^{\circ}$ (Biretta et al. 1995) or
$30^{\circ}$ (e.g., Bicknell \& Begelman 1996) of the M87 jet.
As for $E_{\rm m,s}$, the value of ${\tilde\rho}$ can presumably take
$\sim 10$ at most (Owen et al. 1990).
In Figure~2, for $R=55~{\rm pc}$ at the brightest knot, knot~A
(Biretta et al. 1991), given ${\tilde\rho}$, $L(\theta)$, and $B$,
now we plot $E_{\rm m}|_{Z=26}$ as a function of $U$
in units of the speed of light $c$.
It is found that even for a conservative value of ${\tilde\rho}=1$
(Figs.~2{\it b}$-$2{\it e}), the particle acceleration is, in the region of
$U<0.42$ for $L(42.5^{\circ})\approx 1.4~{\rm kpc}$, not limited spatially,
but temporally, whereas in the region of $U\geq 0.42$, vice versa.
For a convincing value of $B=640~{\rm\mu G}$ (Fig.~2{\it c};
Meisenheimer et al. 1996), which is in a plausible range of
$B=310~{\rm\mu G}$ to $4.59~{\rm mG}$ (Figs.~2{\it b}$-$2{\it d};
{\it hatched area}; Owen et al. 1989),
$E_{\rm m}|_{Z=26}\sim 1~{\rm ZeV}$ is fairly achieved for $U\geq 0.42$,
although in a marginal value of $B=28~{\rm\mu G}$ (Fig.~2{\it e};
Heinz \& Begelman 1997), it reduces to $\sim 50~{\rm EeV}$.
In an optimistic parameter set of ${\tilde\rho}=2$,
$L(30^{\circ})\approx 1.8~{\rm kpc}$, and $B=4.59~{\rm mG}$
(Fig.~2{\it a}), $E_{\rm m}|_{Z=26}\sim 10~{\rm ZeV}$ is achievable
for $U\sim 0.5$, whereby $E_{\rm m}|_{Z=1}\sim 400~{\rm EeV}$.

The validity of this DSA model could be corroborated by the
complementary calculation of the maximum energy of an accelerated electron.
For an electron with energy much lower than $E_{\rm m}$,
an eminent increase of acceleration efficiency is expected,
on account of $\kappa_{\rm n}/\kappa_{\rm n,MHD}\ll 1$.
In regard to the energy limit, it is sufficient to consider the
synchrotron loss (for M87; Biermann \& Strittmatter 1987) with its
timescale of $t_{\rm syn}\propto\gamma^{-1}B^{-2}$, where
$\gamma$ is the Lorentz factor of the electron.
Taking the balance of $t_{\rm acc}$ for $Z=1$ with $t_{\rm syn}$
yields the expression of the maximum Lorentz factor, which scales as
$\gamma_{\rm m}\propto R^{1/3}U^{2/3}$.
A noticeable thing here is that both $t_{\rm acc}$ and $t_{\rm syn}$ have a
common dependence of being proportional to $B^{-2}$, so that $B$ dependence of
$\gamma_{\rm m}$ disappears, in contrast to the expression proposed by
Biermann \& Strittmatter (1987).
The resultant maximum Lorentz factor $\gamma_{\rm m}\sim 10^{11}$, predicted
for $R=55~{\rm pc}$ and $U\sim 0.5$, is amenable to the recent
observational results that indicate no steep synchrotron cutoff
even in the X-ray band of $10^{17}-10^{18}~{\rm Hz}$ (reflecting
$\gamma\sim 10^{7}-10^{8}$: Marshall et al. 2002; Wilson \& Yang 2002),
considerably higher than the previously suggested cutoffs of
$\sim 10^{15}~{\rm Hz}$ (e.g., Meisenheimer et al. 1996;
Heinz \& Begelman 1997).

\section{CONCLUSIONS}

We have constructed a scenario of strong, off-resonant scattering of test
particles by the quasi-static magnetized filaments created via the current
filamentation instability, to model the DSA in AGN jets.
For the M87 jet as an example, considering the most severe energy
restriction from the spatiotemporal scale, we have estimated the
achievable highest energy of accelerated particles at knot~A.
The results indicate that there is a wide range of plausible parameters
where a shock can energize an iron nucleus to $\sim 1~{\rm ZeV}$
(whereby a proton to $\sim 40~{\rm EeV}$) and more.
At the moment, the problem of intergalactic transport of particles
remains unsolved, though M87 is nearby (about $16~{\rm Mpc}$ from us),
so that the effects of a collision with the microwave background are not so
significant (Stecker \& Salamon 1999).
The present consequences might provide a key to elucidate the origin of
EHE cosmic rays with energy above $100~{\rm EeV}$,
in the context of the point-source model.\\

\end{document}